\documentclass{pasj02} 
\usepackage[switch,mathlines]{lineno} 
\usepackage{newtxtext,newtxmath} 
\usepackage{savesym}
\savesymbol{square}
\usepackage{ulem}
 \usepackage{amsmath,amsfonts}
 \usepackage{bm}
 \usepackage{graphics}
 \usepackage{graphicx}
 \usepackage{here} 
 \usepackage{type1cm}
 \usepackage{float}
\usepackage{wrapfig}
\usepackage{multirow}
\usepackage{booktabs}
\usepackage{subcaption} 
\usepackage{pdfpages}
\usepackage{textpos}
\usepackage{color}
\usepackage{url}
\jyear{2026}
\Received{}
\Accepted{}

\begin{document} 

\title{New assignments of the CaH $A^2\Pi-X^2\Sigma^+$ transitions in the sunspot umbral spectrum and effective temperature estimation using molecular lines}

\author{Kosei \textsc{Usami},\altaffilmark{1,2,3}\altemailmark\orcid{0009-0004-9671-7301}\email{kousei.usami@grad.nao.ac.jp, kokokouseiusami@gmail.com} 
 Kaori \textsc{Kobayashi},\altaffilmark{4}\altemailmark\orcid{0000-0002-4163-9498} \email{kaori@sci.u-toyama.ac.jp} 
 Hiroyuki \textsc{Maehara},\altaffilmark{5,6}\altemailmark\orcid{0000-0003-0332-0811}\email{hiroyuki.maehara@nao.ac.jp}
 }
\altaffiltext{1}{Department of Astronomical Science, The Graduate University for Advanced Studies, SOKENDAI, 2-21-1 Osawa, Mitaka, Tokyo, 181-8588, Japan}
\altaffiltext{2}{National Astronomical Observatory of Japan, 2-21-1 Osawa, Mitaka, Tokyo, 181-8588, Japan}
\altaffiltext{3}{Astrobiology Center, 2-21-1 Osawa, Mitaka, Tokyo, 181-8588, Japan}
\altaffiltext{4}{Faculty of Science, University of Toyama, 3190 Gofuku, Toyama 930-855, Japan}
\altaffiltext{5}{Okayama Branch Office, Subaru Telescope, National Astronomical Observatory of Japan, NINS, 3037-5 Honjo, Kamogata, Asakuchi,
Okayama 719-0232, Japan}
\altaffiltext{6}{Astronomical Observatory, Kyoto University, Kitashirakawa-Oiwake-cho, Sakyo, Kyoto 606-8502, Japan}



\maketitle

\begin{abstract}
High-resolution spectroscopy observations of sunspots offer a unique natural laboratory for detailed molecular spectroscopy. Calcium monohydride (CaH) is a vital spectroscopic tracer in cool stellar and solar environments, where its electronic transitions are used for line identification and temperature diagnostics, yet its high-excitation transitions remain poorly characterized. Using high-resolution sunspot umbral spectra obtained with the Brault Fourier-transform spectrometer at the McMath–Pierce telescope \citep{Wallace1999AnAO}, we investigated the $A^2\Pi–X^2\Sigma^+$ electronic transition of CaH in the $\rm{14400-14900\;cm^{-1}}$ region. We report the assignment of 224 spectral lines spanning vibrational bands with $v^{\prime}, v^{\prime\prime} = 0\text{-}3$; notably, 75 lines in the (3-3) band are reported here for the first time, and the identification of the remaining bands was extended to significantly higher rotational quantum numbers ($J\rm{ _{max}\approx 49.5}$) compared to previous studies.
Incorporating these new identifications into spectral simulations that account for overlapping TiO features, we aimed to reproduce the observed umbral spectrum and found modest but measurable improvement in the agreement between the simulated and observed spectra.
Minimizing the residuals between observed and simulated spectra, we estimated an effective umbral temperature range of around 4000 K. We compare this effective temperature against those inferred from independent methods to show that molecular features primarily form in the cool umbral cores of the sunspots. This study demonstrates the utility of solar observations in studying the high-energy transitions that are challenging to reproduce in laboratory settings and the potential of molecular lines as sensitive "thermometers" for cool astronomical objects.
\end{abstract}


\section{Introduction}
The CaH (calcium monohydride) molecule has been observed in the optical spectra of cool stellar atmospheres, such as sunspots and M dwarfs \citep{Olmsted1908,Engvold1973,Reid1997,Wallace1999AnAO,Wallace2000AnAO} primarily through its $A^2\Pi\; - \; X^2\Sigma^+$ and/or $B^2\Sigma-X^2\Sigma$ electronic transitions. The strength of these transitions serves as a critical indicator for the spectral classification of low-mass stars and subdwarfs. This classification provides essential estimates for their effective temperatures and metallicities, which are necessary for accurately placing these objects on the Hertzsprung-Russell (HR) diagram \citep{Barbuy1993,Gizis1997,Kirkpatrick2005,Jao2008,Lepine2008}.
Although not yet detected, searches for CaH in the molecular clouds have been conducted using a radio telescope \citep{Sakamoto1998}. 
Recently, CaH has attracted significant interest within the field of exoplanetary science. With recent advancements in atmospheric characterization techniques, 
it is one of the species expected to be observed in the atmospheres of hot Jupiter (e.g., \cite{Sedaghati2017,Johnson_2023,Petz_2025}).
Comprehensive line lists and pressure-dependent cross sections for this molecule, aimed at providing more accurate input opacities for radiative transfer models have been developed \citep{Tennyson_2016, Owens2022}.
The anticipated presence of CaH in exoplanetary atmospheres is motivated by the case of TiO (titanium monoxide). 
Since TiO and CaH are observed under similar temperature conditions in sunspots or late type star\citep{Richardson1931,Engvold1973,1976A&AS...24..111B,Wallace1999AnAO,Lepine2008}, the successful detection of TiO in several hot Jupiters \citep{Nugroho2017, Yan2019, Cont2021, Ouyang2023} may suggest that CaH could also be an identifiable species in these high temperature environments. 

Previous studies have indicated that the choice or quality of molecular line lists including CaH can influence the interpretation of atmospheric observables \citep{Gharib-Nezhad2021}, highlighting the ongoing importance of refined molecular data including accurate transition frequencies and Einstein A coefficients, in both stellar and planetary contexts. Pressure-broadening coefficients are also relevant for line-shape calculations, although such data remain limited for CaH. Therefore, spectroscopic studies-specifically those improving the precision of CaH line lists, are essential for accurately estimating the temperature, luminosity, metallicity and elemental abundances in stellar photospheres, even molecule detection in exoplanetary atmospheres.

In molecular spectroscopy, CaH has long been a fundamental subject of research since its detection in sunspots \citep{Olmsted1908}.
As a simple diatomic molecule, its electronic states and vibrational and rotational levels provide an ideal system for benchmarking quantum chemical calculations.
In particular, the electronic excited state $B/B^2\Sigma^+$\citep{SHAYESTEH201346,GDBell1979,Watanabe2016,Watanabe2018} is known to have a double minimum potential. To understand these energy states, research is actively being conducted in laboratory molecule spectroscopy. 
In addition to the electronic excited state $B/B^2\Sigma^+$, studies on other electronic states ($X^2\Sigma^+, A^2\Pi, 1^2\Delta, E^2\Pi, D^2\Sigma^+$) range from rotational states to electronic states \citep{SHAYESTEH201346,RAM201186,GDBell1979,Furuta2025a,Furuta2025}.

While laboratory spectroscopy is vital, measuring transitions involving high-energy states or low transition probabilities is often challenging due to limitations in temperature and/or particle density under laboratory conditions. However, these transitions could be detected by utilizing spectra obtained under high-temperature astronomical conditions, such as those in the solar photosphere and sunspots, where high-precision observations are feasible. Information regarding high-energy states facilitate the refinement of molecular constants and the revision of theoretical predictions. Specifically, observations of sunspot umbra using the McMath-Pierce telescope at Kitt Peak reported the assignment of rotationally-resolved lines in the CaH $A^2\Pi–X^2\Sigma^+$ spectrum corresponding to the (0-0), (1-1), and (2-2) vibrational bands \citep{Wallace1999AnAO}. Recent advancements in laboratory spectroscopy now allow us to consider higher temperatures and anticipate the observation of higher-energy levels, including the (3-3) vibrational band. Moreover, sunspot umbral spectra are observed at high resolution, enabling the extension of identifications to high-lying transitions that are difficult to obtain in the laboratory.

In this study, we leverage high-resolution sunspot umbral spectra as "natural laboratory" to explore the high excitation transitions of CaH that were not assigned in the previous study \citep{Wallace1999AnAO}. Our primary objective is to provide a comprehensive assignment of the $A^2\Pi–X^2\Sigma^+$ system, specifically focusing on the first-ever identification of the (3-3) vibrational band and the extension of (0-0), (1-1), and (2-2) bands to high rotational quantum numbers. Furthermore, we incorporate these new CaH identifications into a spectral simulation that accounts for overlapping TiO features. This approach allows us to achieve a high-precision reproduction of the sunspot umbral spectrum and estimate the effective umbral temperature; Our results suggest an effective temperature around 4000 K, which suggests that the molecular features are sensitive to low temperature regions within the sunspot temperature range. These results demonstrate the utility of solar or high-resolution stellar observations in improving the spectroscopic data required for characterizing cool stellar and exoplanetary atmospheres.

\section{Data and Method}

\footnotetext[1]{https://nispdata.nso.edu/ftp/pub/atlas/spot3atl/}
\footnotetext[2]{PGOPHER version 10.0, C M Western, 2017, University of Bristol Research Data Repository, doi:10.5523/bris.160i6ixoo4kir1jxvawfws047m.}
\footnotetext[3]{http://www.phys.
appstate.edu/spectrum/spectrum.html}

\subsection{Sunspot Umbral Spectra Data}\label{sec:2}
We have chosen the sunspot umbral data observed by the Brault Fourier transform spectrometer mounted in McMath-Pierce Solar Telescope, Kitt Peak (1981/03/24 $\#$1 by L. Testerman). The spectrum was taken from the NSO website\footnotemark[1] \citep{Wallace1999AnAO}. The sunspot umbral spectra used in this study cover the red region. We used spectra spanning the ranges of $\mathrm{14000\;cm^{-1}-15000\;cm^{-1}}$,  which had already been corrected for telluric lines. Previous studies have also reported various molecular spectra for the same spectral data \citep{Wallace1999AnAO}.
In the region addressed in this study, the CaH ($A^2\Pi – X^2\Sigma^+$) band and TiO ($A^3\Phi–X^3\Delta$,$\mathrm{\gamma}$) band are particularly strong. TiO lines were also identified (e.g., \cite{Fowler1907,Engvold1973,Sotirovski1971,1976A&AS...24..111B,Berdyugina2003}).

\subsection{Molecular Data}
In this study, spectral simulations of the $A^2\Pi–X^2\Sigma^+$ electronic transition in CaH molecules were performed using the simulation and fitting software PGOPHER\footnotemark[2]\citep{WESTERN2017221}.
PGOPHER outputs detailed transitions list.
The molecular constants for CaH used here were adopted from the laboratory Fourier-transform emission study of the $A^2\Pi-X^2\Sigma^+$ \citep{SHAYESTEH201346}.  
The Einstein A coefficients of each band were adopted to simulate intensities from the quantum chemical calculation \citep{SHAYESTEH2017345}. 
Therefore, the conversion of the Einstein A coefficient for electronic transition A in the CaH molecule to $S_{J' \rightarrow J''}$ was performed using equation (5.143) from \citep{bernath2016spectra}, shown below.

\begin{equation}
 S_{v'\rightarrow v''}=\frac{3.1885834\times10^6}{\tilde{\nu}^3}A_{v'\rightarrow v''}
\end{equation}
We used the square-root of this formula for the band from state $v'$ to state $v''$ as the strength in the PGOPHER, generating the rotationally-resolved line list with oscillator strength.

The Herman-Wallis effect, affecting the relative intensities of the branches, was reported in the CaH electronic transition  \citep{10.1093/mnras/stx2681}. The new molecular constants have little effect on the line intensity. The $J$-dependent terms of the dipole transition moments to reproduce Herman-Wallis effect were not directly provided in the paper.  The highest $J$ reported in their supplement was not high enough for our study and this effect was not considered. 

\subsection{Simulation of Sunspot Umbral Spectra}
We generated line lists from the PGOPHER simulation and 
the molecular constants of CaH and TiO used for the PGOPHER input were obtained from \citep{SHAYESTEH2017345,Ram1999} respectively.  The set of line list was modified to the appropriate format and transferred to the astronomical simulation. The simulation of sunspot umbral spectra was performed by using software SPECTRUM\footnotemark[3]\citep{1994AJ....107..742G} in the $\mathrm{14000\;cm^{-1}-15000\;cm^{-1}}$ region. In this simulation, CaH and TiO were assumed as the primary molecular species, focusing on the electronic transition from v=0 to v=3, and  from v=0 to v=4, respectively, due to their dominant contribution to the umbral opacity in this wavenumber range \citep{Wallace1999AnAO}. To evaluate the validity of this approach,
We have imported additional molecular and atomic lines listed in the cool7.iso.lst (private communication:  Dr. Richard Gray).
This list includes lines from $\text{TiO}$ (44\%), $\text{C}_2$ (30\%), $\text{CN}$ (20\%), $\text{CaH}$ (5\%) and other atoms and molecules (1\%).  The numbers in the parentheses are the fraction of lines ($\text{lines}_{\text{species}}/\text{lines}_\text{total}$).
Hydride molecules usually have large rotational constant and vibrational frequencies, resulting in fewer lines. However, our tests demonstrated that the inclusion of $\text{C}_2$ and $\text{CN}$ has a negligible impact on the synthetic spectrum globally and the final results (See Appendix 1, 2).
This confirms that the systematic deficiency in line opacity from other species is insignificant for the objectives of this study.
We used the model atmosphere for the solar metallicity ([M/H]=0) and log g=4.5 several effective temperatures ranging from 3500 K to 4500 K ($T_{\text{eff}}(K) = 3500, 3750, 4000, 4250, 4500 $) taken from "ap00k0odfnew.dat" on Kurucz's Grid of model atmosphere\footnotemark[4] \citep{Castell_Kurucz}.
We calculated the synthetic spectra assuming a microturbulence of $\xi_{\mathrm{t}} = 2.0 \mathrm{~km~s^{-1}}$. Both macroturbulence ($v_{\mathrm{macro}}$) and rotational velocity ($v \sin i$) were assumed to be zero, considering the suppression of convective motion by the strong magnetic field in the sunspot umbra and the spatially resolved nature of the observations. We adopted the solar chemical abundances from \citet{Asplund2021}, a widely used recent reference, specifically $\log \epsilon(\mathrm{Ca}) = 6.30 \pm 0.03$ and $\log \epsilon(\mathrm{Ti}) = 4.97 \pm 0.05$ 
(on the scale where $\log \epsilon(\mathrm{H}) = 12$). Hereafter, these are abbreviated as A21. In Appendix 2, we investigated the sensitivity of the effective temperature estimates to the adopted abundance set by comparing \citep{Gaur1972, Grevesse1996, Asplund2021, Magg2022}. It is shown that the three most recent abundance sets give consistent results, while the values from \citep{Gaur1972} also lead to similar estimates. Finally, the model spectra were convolved with a Gaussian profile to match the instrumental resolution of $R = 600,000$ \citep{Wallace_2011} using the SMOOTH2 within the SPECTRUM \citep{1994AJ....107..742G}. 
\footnotetext[4]{http://kurucz.harvard.edu/grids.html}

\section{Result}
\subsection{Assignment of Sunspot Umbral Spectra}
\begin{table}[h]
 \tbl{Comparison of the number of assigned lines in sunspot umbra}{%
  \begin{tabular}{ccc} \hline\hline
    & this work$^*$ & Wallace et al. (1999) \\ \hline
   0-0 & 195 (45) & 150  \\ 
   1-1 & 177 (29) & 148 \\ 
   2-2& 146 (75) & 71 \\ 
   3-3& 75 (75) & 0 \\ \hline
  \end{tabular}}\label{table1}
  \begin{tabnote}
  \footnotemark[*]{() indicates newly assigned lines in this work.} \\ 
  \end{tabnote}
\end{table}
\begin{table}[h]
 \tbl{Comparison of $J_{\mathrm{max}}$}{%
  \begin{tabular}{cccc} \hline\hline
  & sunspot (this work) & sunspot$^1$ & laboratory$^2$ \\ \hline 
  0-0 & 49.5 & 47.5 & 53.5\\ 
   1-1 & 49.5 & 43.5 & 45.5\\ 
   2-2 & 49.5 & 32.5 & 38.5\\ 
   3-3 & 30.5 & \- & 26.5\\ \hline
  \end{tabular}}\label{table2} 
  \begin{tabnote}
  \footnotemark[1]{ \citet{Wallace1999AnAO} } \\ 
  \footnotemark[2]{ \citet{SHAYESTEH2017345} } \\ 
  \end{tabnote}
\end{table}
In this study, assignments were done within the range from $\mathrm{14400\;cm^{-1}}$ to $\mathrm{14900\;cm^{-1}}$.
Table 1 shows a comparison of the number of assigned lines for CaH electronic transitions in this study and in previous study. 
Table 2 compares the highest rotational quantum numbers with those in the previous studies of sunspot \citep{Wallace1999AnAO}, and laboratory spectroscopy \citep{SHAYESTEH2017345}.
The assignments were extended to the higher rotational quantum numbers, and 75 lines of (3-3) were assigned for the first time in this study.
Details of each transition are given on supplementary.
\subsection{Reproduction of Sunspot Umbral Spectra}
\begin{figure}[H]
 \begin{center}
 \includegraphics[width=8cm]{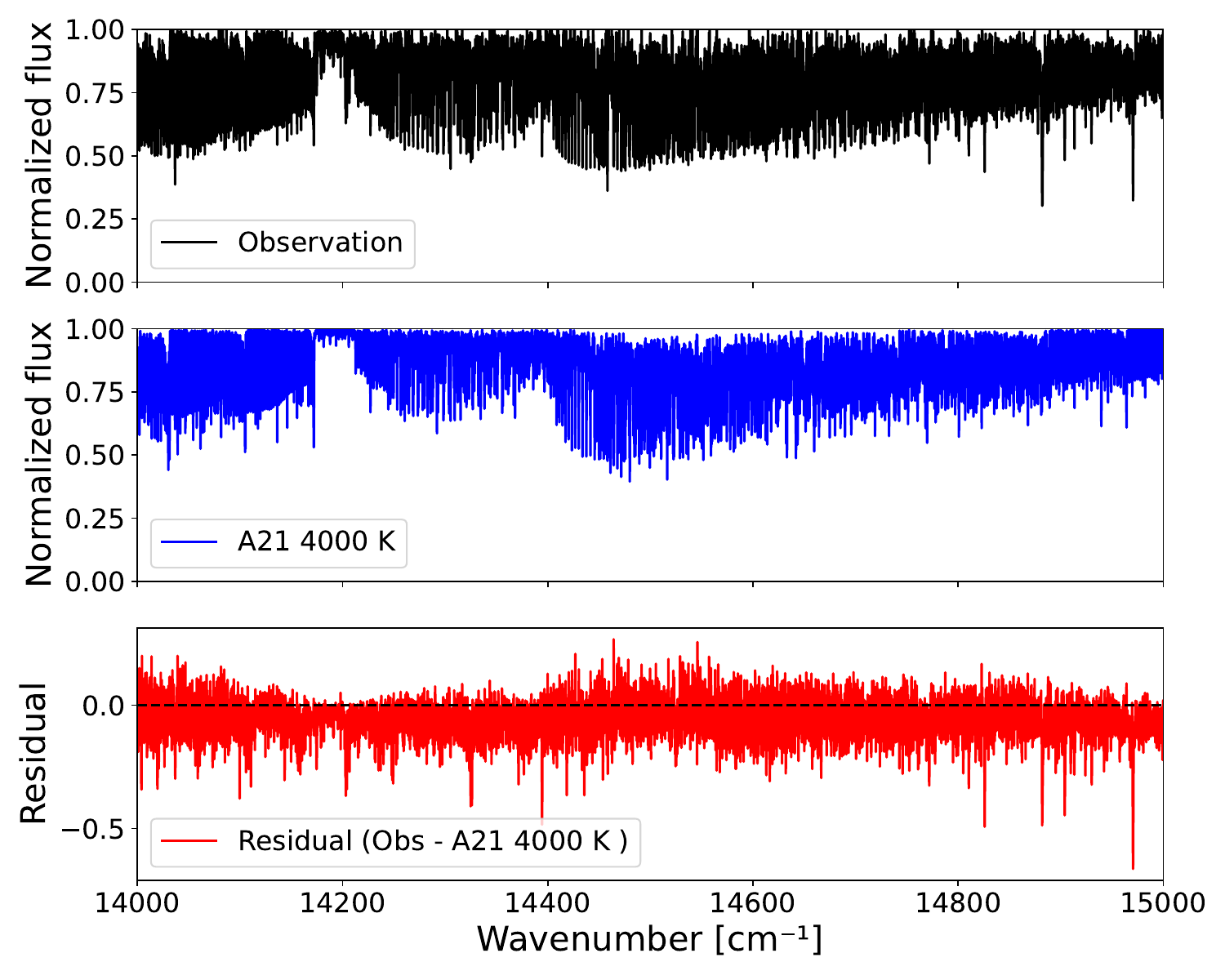} 
 \end{center}
\caption{The comparison between observation and simulation spectra. The upper middle, bottom panels show observation, the most representative simulation by using line list $\#$1 assuming \citep{Asplund2021} at 4000 K,and the difference (obs-calc).\\ {Alt text: Three vertically stacked panels showing spectra. The top panel is the observed spectrum, the middle panel is the simulated spectrum at 4000 Kelvin, and the bottom panel shows the difference between the observed and simulated spectra.}}
\label{fig}
\end{figure}
Based on the assignment of the newly observed (3-3) transition and the higher $J$ transitions of (0-0), (1-1) and (2-2), we investigated the extent to which these lines contribute to the spectral simulation and whether we need to consider higher energy transitions. In principle, the intensity of higher energy transitions is weaker than that of lower energy transitions due to various factors, including
transition probability and level populations.
Therefore, we generated two line lists to quantify the significance of these high-energy lines. The first (line list $\#$1) includes transitions of $J_\mathrm{max}$ up to 99.5 in all vibrational levels (v = 0–3). The second set (line list $\#$2) includes only transitions of $J_\mathrm{max}$ up to the values of those  shown in Table 2, i.e., the previously assigned transitions in the sunspot umbra \citep{Wallace1999AnAO}. The (3-3) lines are not included in this line list $\#$2.
Here, regarding TiO transitions, both line lists included electronic transitions from v = 0 to v = 4 \citep{Ram1999}.  
Figure 1 compares the observed umbral spectrum with the simulated spectrum by using line list $\#$1 that best represents the data — i.e., the simulation minimizing the sum of squared residuals over the explored temperature range. The upper panels show the observed spectrum, and the middle panels show the corresponding simulated spectrum over the same wavenumber region. The figures of other conditions are shown in figures in the Appendix 3. Some of the strong features of the residuals include the known atomic lines such as Fe, Ti or Ni \citep{Wallace1999AnAO}.
In the calculation of the sum of squared residuals, regions corresponding to strong atomic lines were masked using a specific wavenumber window (center ± 0.3 cm$^{-1}$).
\\Figure 2 presents the sum of the squared residuals of the intensity between the observed and simulated spectra as a function of temperature from 3500 K to 4500 K. 
Filled and open star and diamond symbols indicate the  line list $\#$1 and the line list $\#$2 respectively. Our simulations indicate best reproduction at temperatures of 4000 K for A21 with line list $\#$1. To account for the uncertainties in the reference solar abundances, error bars in Figure 2 were calculated by varying the Ca and Ti abundances within their reported $1\sigma$ uncertainties \citep{Asplund2021}. Note that in Figure 2, line list $\#$2 have smaller sum of squared residuals in A21 3500 K and 3750 K than the line list $\#$1 ($J_\mathrm{max}$ up to 99.5). 
This may be because simulated absorption lines  become stronger at lower temperature.
See Fig 6 and 7 to check the reproduction using line list $\#$1 at different temperatures.
\begin{figure}[H]
 \begin{center}
 \includegraphics[width=8cm]{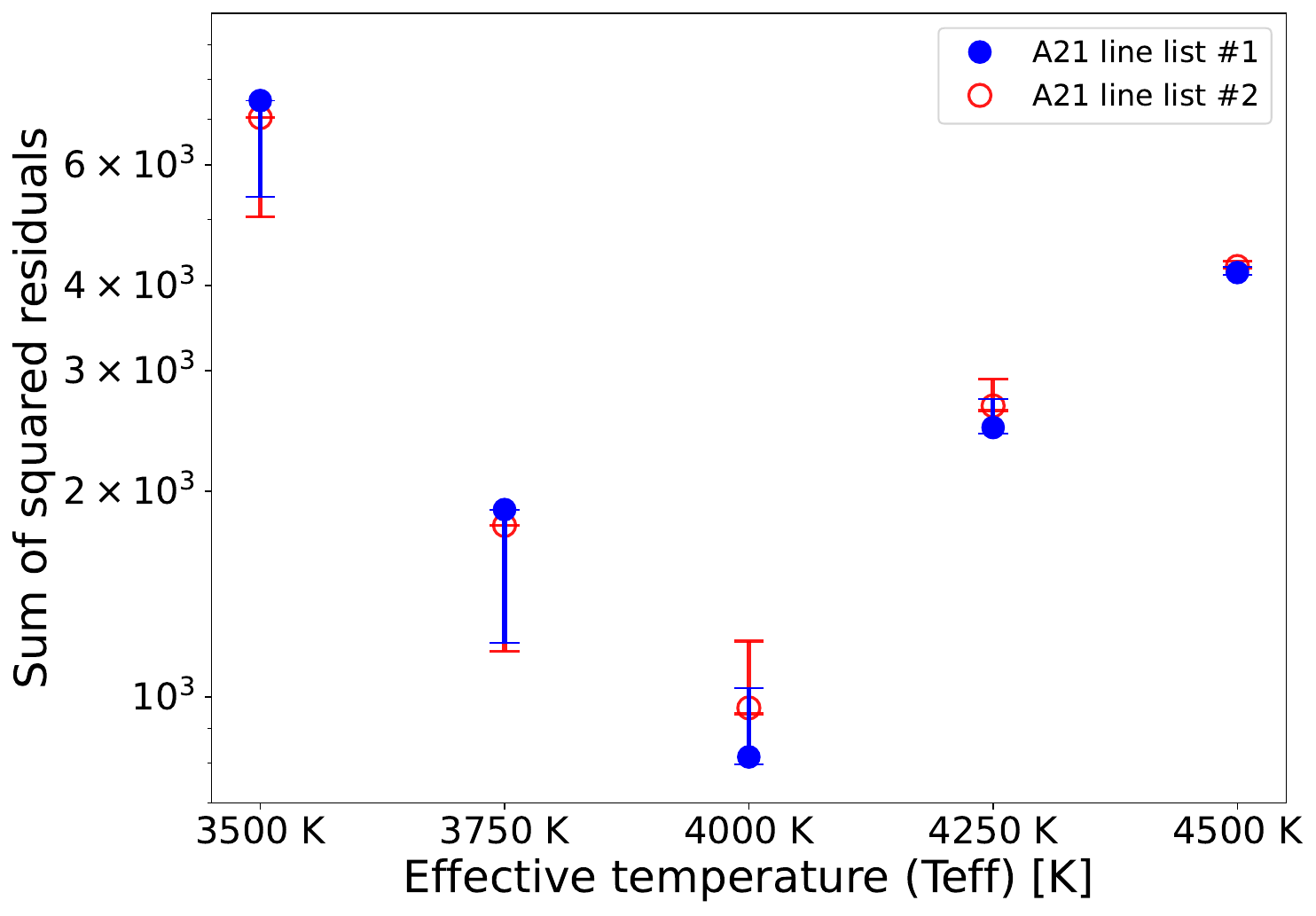} 
 \end{center}
 \caption{Temperature versus the sum of squared residuals for assumed abundances A21 (\cite{Asplund2021}). Filled circle symbols represent the sum of squared residuals obtained using the line list $\#$1 for abundance A21, while open circle
 symbols indicate results from the line list $\#$2 (Section 3.2). The error bars represent the range of the sum of squared residuals resulting from the $1\sigma$ uncertainties in the solar chemical abundances of Ca and Ti as reported by \citet{Asplund2021}.} Results are shown for temperatures of 3500, 3750, 4000, 4250, and 4500 K.\\{Alt text: Scatter plot showing how the sum of squared residuals changes with temperature from 3500 to 4500 Kelvin. The data compares the results for the abundances, A21, across two different line lists, number 1 and number 2. It demonstrates the sensitivity of the residuals to the choice of line list for each temperature point.}
\label{fig2}
\end{figure}
Figure 3 shows part of the simulation for different line lists $\#$1 and $\#$2 with T = 4000 K and the abundance A21. The red dashed lines represent newly assigned (3-3) line positions. Comparing the two line lists shows that simulations using the line list $\#$1 that incorporates the new transition better reproduce the observed black sun spot spectrum.
\begin{figure}[H]
 \begin{center}
 \includegraphics[width=8cm]{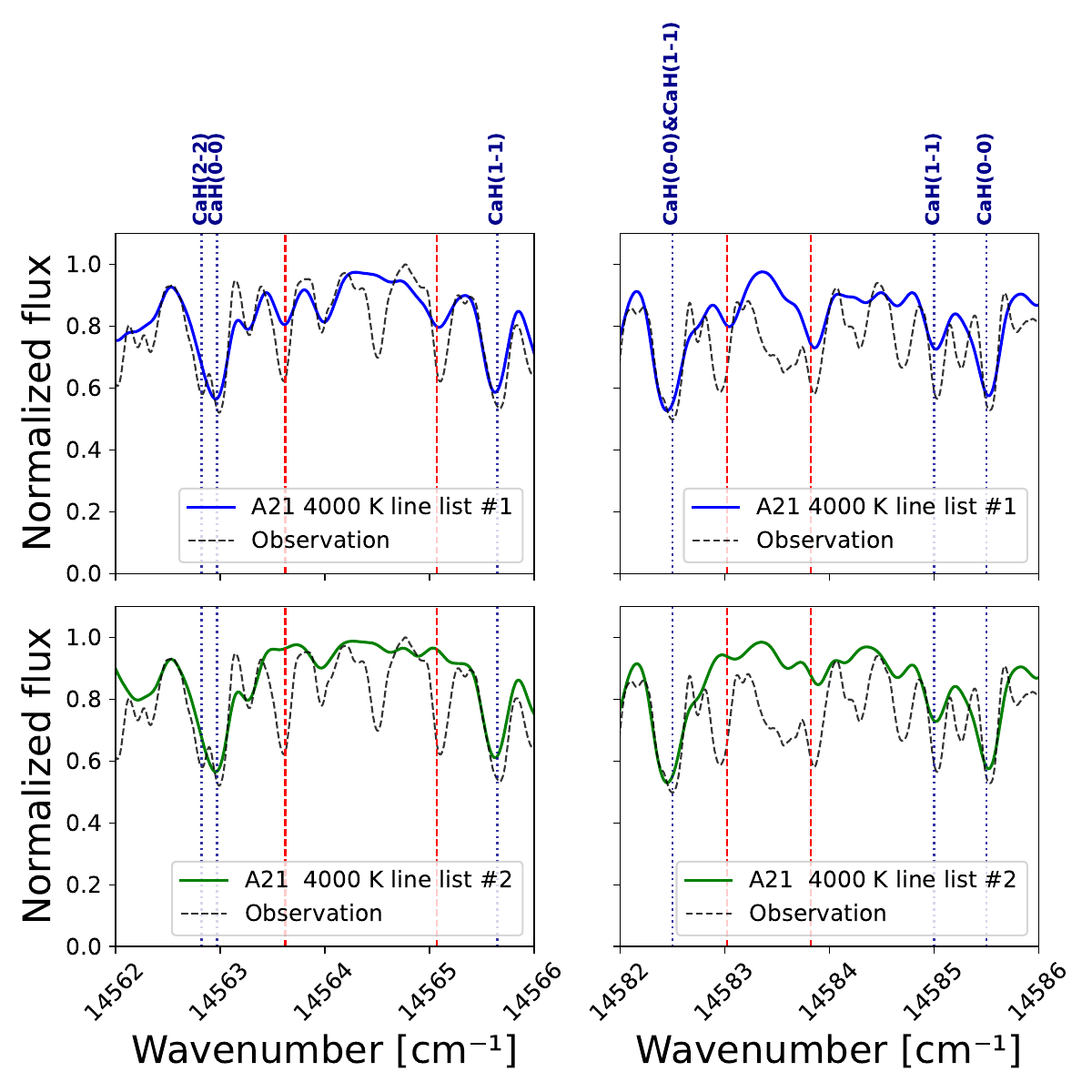} 
 \end{center}
\caption{Examples of newly assigned (3-3) line contribution. Each upper and lower panels show simulation results with different line lists $\#$1 and $\#$2.
Red dashed line indicates newly (3-3) lines and previously assigned lines show as blue dotted ones. 
The difference between the line lists are same as those in Figure 2 and mentioned in Section 3.2.\\{Alt text: Comparison of line contribution simulations shown in two panels. The upper panel shows results calculated with line list number 1 and the lower panel shows results with line list number 2. Both panels compare the spectral contribution of newly assigned three dash three lines with the contribution of previously assigned lines.} }
\label{fig3}
\end{figure}
\section{Discussion}
\begin{longtable}{lccc}
  \caption{Brightness-Size Empirical Relation Parameters} \label{tab:model_params}
\hline\noalign{\vskip3pt} 
  Empirical relation & Parameter & Values & Equation \\ [2pt]
\hline\noalign{\vskip3pt} 
\endfirsthead      
\hline\noalign{\vskip3pt} 
  Empirical relation & Parameter & Values & Equation \\ [2pt]
\hline\noalign{\vskip3pt} 
\endhead
\hline\noalign{\vskip3pt} 
\endfoot
\hline\noalign{\vskip3pt} 
\multicolumn{4}{@{}l@{}}{\hbox to0pt{\parbox{160mm}{\footnotesize
\hangindent6pt\noindent
\hbox to6pt{\footnotemark[]\hss}\unskip%
This table summarizes the empirical relations used to estimate sunspot umbral brightness from their size, \\based on previous studies. The table lists the fitting parameters and corresponding equation, including \\both power-law and linear models. The parameters are taken from \citet{Mathew2007} (M07) \\and \citet{Kiess2014} (K14). These relations describe how the relative intensity \\(core, mean, or minimum) of umbral varies as a function of their radius.
}\hss}} 
\endlastfoot 
 M07 (core)   & $C_{\text{core}}$     & 1.8598  & \multirow{2}{*}{ $ I_{\text{core}} (r_{\text{arcsec}}) = C_{\text{core}} \cdot r_{\text{arcsec}}^{\alpha_{\text{core}}}$ } \\
 (power-law)  & $\alpha_{\text{core}}$ & -1.0679 & \\ \hline
 M07 (mean)   & $C_{\text{mean}}$     & 0.8297  & \multirow{2}{*}{ $ I_{\text{mean}} (r_{\text{arcsec}}) = C_{\text{mean}} \cdot r_{\text{arcsec}}^{\alpha_{\text{mean}}}$ } \\
 (power-law)  & $\alpha_{\text{mean}}$ & -0.3052 & \\ \hline
 M07 (core)   & $A_{\text{core}}$     & -0.0552 & \multirow{2}{*}{ $I_{\text{linear}} (r_{\text{arcsec}}) = A_{\text{core}} \cdot r_{\text{arcsec}} + B_{\text{core}}$} \\
 (linear)        & $B_{\text{core}}$     & 0.6515  & \\ \hline
 K14 (min)     & $C_{\text{min}}$      & 0.830   & \multirow{2}{*}{$ I_{\text{min}} (r_{\text{Mm}}) = C_{\text{min}} \cdot r_{\text{Mm}}^{\alpha_{\text{min}}} $} \\
 (power-law)   & $\alpha_{\text{min}}$  & -0.958  & \\
\end{longtable}
\subsection{Effective Temperature of a Sunspot Umbral and Comparison with Previous Brightness Temperature Studies}
In this study, the effective temperature of a sunspot umbra was estimated by comparing observed sunspot umbral spectra with simulated ones. Figure 2 shows the sum of squared residuals for the simulations and observations calculated at each temperature. As the figure shows, the analysis in this study minimized the sum of squared residuals in the $4000 \text{ K}$, providing the best reproduction of the observed spectra. Several studies of sunspots have revealed umbral temperatures on the order of several thousand Kelvin. Based on separate measurements of umbral brightness (e.g., \cite{Maltby1970,Albregtsen1981,Bray1981}), the effective temperatures of an umbra are suggested as 3900 K - 4800 K \citep{Sloanki2003}. However, simulations with temperatures above 4200 K, showed that the absorption lines from CaH and TiO molecules were significantly weaker than observed(See Appendix 2), suggesting CaH and TiO molecules we observed originates from low temperature regions within sunspot temperature range. 
\subsection{The Effective Rotational Temperature of CaH $\&$ TiO}
Molecular absorption lines visible in sunspot umbral spectra provide a diagnostic of the temperature conditions in the line-forming region where the molecules exist. Since the rotational structure of molecular electronic transitions is sensitive to temperature, an effective rotational temperature can be inferred from the observed line strengths. 
Several research groups have reported rotational temperatures for different molecular species using the sunspot umbral spectra.
\citet{Behere2020} analyzed the $B^2\Sigma-X^2\Sigma$ electronic spectrum of the CaH molecule around the 15000 cm$^{-1}$ - 16000 cm$^{-1}$ region. 
They determined its rotational temperature to be $4164 \pm 164$ K.
Similarly, \citet{Webber1971} reported rotational temperatures of $3520 \pm 300$ K and $3350 \pm 400$ K for the CaH $A^2\Pi–X^2\Sigma^+$ (0-0) and (1-1) bands, respectively. 
Our result is therefore broadly consistent with previous CaH based estimates \citep{Webber1971, Behere2020}.
On the other hand, rotational temperatures derived from TiO show a much larger variation, ranging from $2000\;\rm{K}$ to $\;3700\;\rm{K}$ among previous studies (e.g., \citet{Makita1968,Webber1971,Sotirovski1971,Sriramachandran2020}). For example, \citet{Sriramachandran2020} reported a relatively low temperature of $ 2555 \pm 780$ K, while \citet{Webber1971} derived higher values of $3480 \pm 300$ K. 
This spread likely reflects differences in the TiO band system and line selections used in each study, as well as possible variations among individual sunspots, all of which affect the line-forming layers and, consequently, the estimated rotational temperatures. 
In our study, the spectral simulation simultaneously reproduces both CaH and TiO features over a broad wavelength range. While rotational temperatures are typically derived from specific molecular bands or lines, our approach provides an effective temperature that reflects the overall spectral behavior.
The resulting temperature of 4000 K is therefore higher than TiO-only rotational temperatures, but this does not indicate an inconsistency, rather it suggests that TiO and CaH are sensitive to different parts of the vertical temperature structure of the sunspot atmosphere.
More generally, previous studies of other molecular species, such as MgH, AlO, FeH, CrH, and VO have also reported distinct rotational temperatures, further supporting the interpretation that each molecule traces a different height range in the sunspot atmosphere \citep{Webber1971,Wallace1999b,Sriramachandran2013,Fawzy2009,Sriramachandran2011,Sriramachandran2008}. These molecular rotational temperatures are also lower than the brightness temperatures typically inferred for sunspot umbrae, suggesting that molecular diagnostics are expected to preferentially sensitive to the cooler components of the atmosphere.
\subsection{Temperature Estimation from Sunspot Umbral Size}
\citet{Collados1994} found that the umbral temperature depends on the size(area), and larger umbrae tends to be cooler. 
Thus, based on empirical relationships from \citep{Mathew2007,Kiess2014}, we attempted to estimate brightness temperature from umbral size. These empirical relations are based on continuum intensity measurements near 617.33 nm ($\rm{16200\;cm^{-1}}$) \citep{Kiess2014} and 676.80 nm ($\rm{14755\;cm^{-1}}$) \citep{Mathew2007}. Table 3 showed each parameters on those relationships.
Here, "core" and "min" denote the minimum umbral intensity value, while "mean" refers to the average intensity value of the umbral. Both empirical relationships are expressed in terms of the relative intensity normalized quiet sun intensity, defined as ($I_{\text{re}}=I_{\rm{umbral}}/I_{\rm{QS}}$). In addition, the definition of relative intensity $I_{\rm{re}}$ differs between \citep{Mathew2007} and \citep{Kiess2014}. \citet{Mathew2007} define the umbral intensity normalized to the quiet Sun at the same heliocentric angle, which is the correction factor for the projection effect $\mathrm{\mu}$, explicitly accounting for center-to-limb variations, 
\citet{Kiess2014}, on the other hand, use the limb-darkening–corrected minimum umbral intensity normalized to the disk-center quiet Sun intensity.
Therefore we used different quiet Sun intensities in each model.
To convert relative intensity to brightness temperature, 
we must first estimate the quiet Sun intensity.
Since no contemporaneous quiet Sun spectrum was available in the data we analyses,
the quiet Sun intensity at each sunspot umbral position was estimated using the empirical solar center-to-limb intensity functions of \citep{Neckel1994,Neckel2003}. 
Specifically, for each observed wavelength $\lambda$ (nm) and projection effect $\mu$ we computed the quiet-Sun specific intensity as
\begin{equation}
  I_{\text{QS}}=I_{\rm{center}}(\lambda)P_5(\mu,\lambda),
  \;P_5(\mu,\lambda)=\Sigma_{i=0}^{5}A_i(\lambda)\mu^i
\end{equation}
where  $I_{\rm{center}}(\mathrm{\lambda})$ and the limb-darkening coefficients 
$\mathrm{A_i(\lambda)}$ from the table of \citep{Neckel2003,Neckel1994}. 
To ensure consistency with the original empirical relations, the quiet-Sun intensity was evaluated at the corresponding wavelengths of each study. For the \citet{Mathew2007} relation, we adopted $\lambda = 669.40\;\mathrm{nm}$ ($\rm{14939\;cm^{-1}}$) and for the \citet{Kiess2014} relation, $\lambda = 610.98\;\mathrm{nm}$ ($\rm{16369\;cm^{-1}}$). These values correspond to the nearest available wavelengths in the tabulated data of \citep{Neckel2003,Neckel1994}.
Note that we used the same $\mathrm{\mu}$ as sunspot umbral size derivation (see Section 4.3.1) in Mathew model.
However, in Kiess model the disk-center value of $\mathrm{\mu=1}$ was applied.
After obtaining $I_{QS}(\mu,\lambda)$, we calculated $T\mathrm{_{\text{QS}}}$.
These values were used as $B_{\lambda}(T_{\text{QS}})$.
Based on the relationships of Table3, relative intensity $I_{\text{re}}$ is determined by the umbral size. Using the $I_{\text{rel}}$, the umbral brightness intensity was calculated as $B_{\lambda}(T_{\text{spot}}) = I_{\text{re}} \cdot B_{\lambda}(T_{\text{QS}})$.
Finally, the brightness temperature of the umbral was calculated by numerically solving the Planck equation for $T_{\text{spot}}$ based on the value of $B_{\lambda}(T_{\text{spot}})$.
\subsubsection{Umbral Size}
\footnotetext[5]{https://solarwww.mtk.nao.ac.jp/en/database.html}
\begin{figure}[h]
  \begin{center}
    \begin{minipage}{0.48\textwidth}
      \centering
      \includegraphics[width=\linewidth]{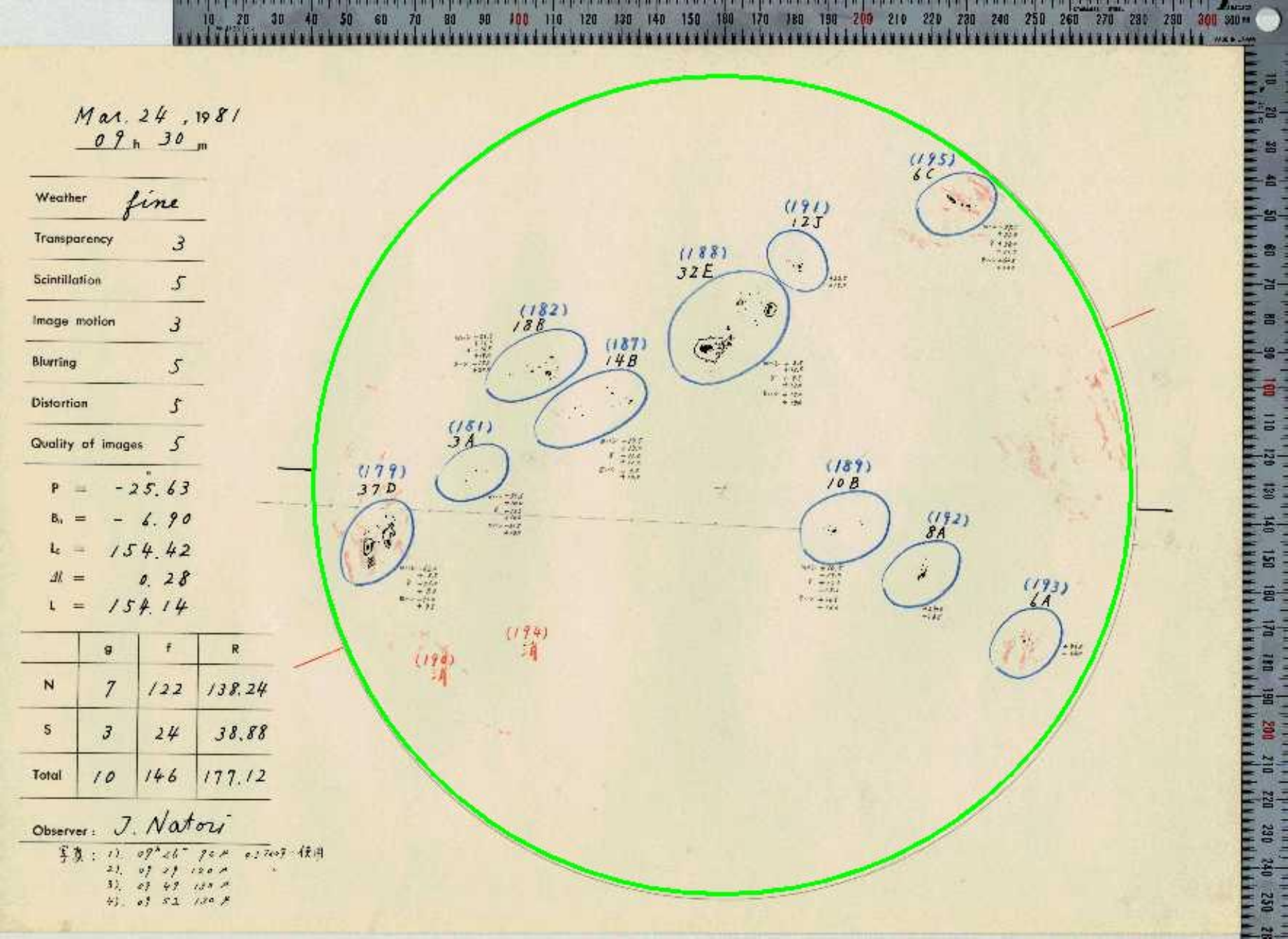}
    \end{minipage}
    \hfill 
    \begin{minipage}{0.48\textwidth}
      \centering
      \includegraphics[width=.55\linewidth]{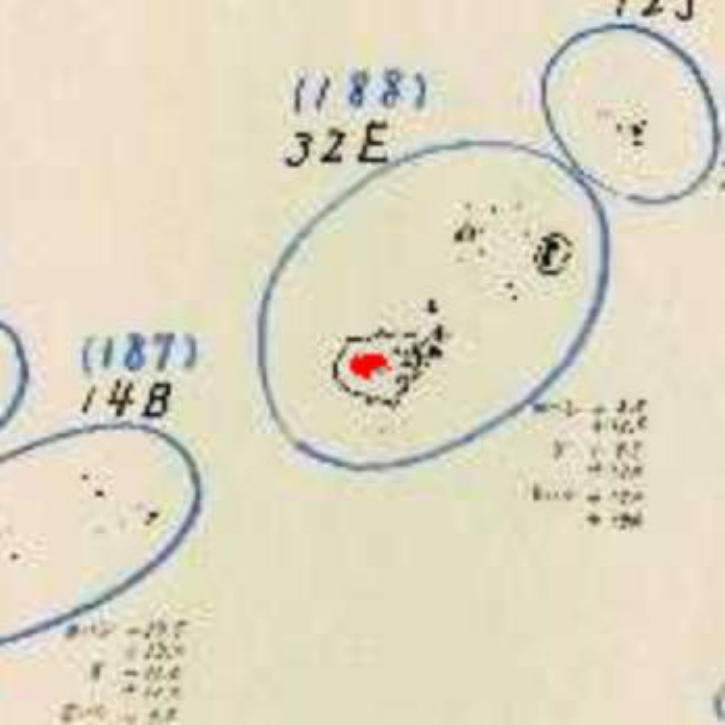}
    \end{minipage}
  \end{center}
  \caption{Sunspot drawing sketch on 1981-03-24, taken from Solar Science Project, National Astronomical Observatory of Japan\protect\footnotemark[5]. The upper panel shows the full-disk sun drawing on 1981-03-24. The lower panel shows a close-up of the sunspot group we analyzed. The green line in the upper panel and red region in the lower panel indicate the solar limb and umbral area extracted with our code.\protect\\ \{Alt text: Sunspot drawing sketch from March 24, 1981, from the National Astronomical Observatory of Japan. The figure consists of two panels. The upper panel is a full-disk drawing of the sun that shows the extracted solar limb. The lower panel is a magnified view of a specific sunspot group that shows the umbral area extracted for the study.\}}
  \label{fig:umbral_area}
\end{figure}
We estimated the umbral size from sunspot drawings by the National Astronomical Observatory of Japan (NAOJ), retrieved from the NAOJ Solar Science Observatory website\footnotemark[5], by extracting the dark area below the threshold in the luminance channel of the image with the umbral detection and area(size) measurement code using the OpenCV library \citep{opencv_library}. Figure 4 shows sunspot drawing images that we used in this work.
We also estimated the center of the sun ($(x_0, y_0)$), the radius ($R_{\text{sun}}$), the center of the sunspot group ($(x_c, y_c)$), and the apparent area of the umbral ($A_{\text{app}}$) in pixels from the sunspot drawing images. The green circle in the top figure of Figure 4 represents the apparent circumference of the Sun. 
To estimate the true umbral area ($A_{\text{true}}$), it is necessary to account for projection effects $\rm{\mu}$. The correction factor for the projection effect ($\mu$) can be calculated by equation (\ref{mu}).
The following equation, $d = \sqrt{(x_c-x_0)^2 + (y_c-y_0)^2}$ yields $\mathrm{\mu}$.
\begin{equation}
  \mu = \sqrt{1 - \left(\frac{d}{R}\right)^2}
  \label{mu}
\end{equation}
Therefore, the true area of the umbral can be estimated as follows.
\begin{equation}
  A_{\text{true}} = \frac{A_{\text{app}}}{\mu}
  \label{eq:projection_correction}
\end{equation}
Those values are calculated as $A\mathrm{_{true}=25.29\;px^2}$ and $\mathrm{\mu=0.9417}$ ($R\mathrm{_{sun}=281.67\;px}$). Then
the effective umbral radius obtained from a circle with the same area as the umbral. We estimated the effective umbral radius using equation (\ref{sunspot_radius})
\begin{equation}
  r = \sqrt{\frac{A_{\text{true}}}{\pi}} \label{sunspot_radius}
\end{equation}
Uncertainties in the sunspot umbral area and effective radius are estimated from the measurement uncertainties in units of pixels.
\subsubsection{Derived Brightness Temperature}
\begin{table}[h]
 \caption{Derived Temperature (K) in Each Empirical Relation}
  \centering
   \begin{tabular}{cccc} \hline\hline
   M07(core) & M07(mean) & M07(core) & K14(min)\\  
   (power-law) & (power-law)  & (linear) & (power-law)\\ \hline
   4047 $\mathrm{\pm 129}$ & 4893 $\mathrm{\pm 21}$& 3797 $\mathrm{\pm 111}$& 4036 $\mathrm{\pm  69}$\\ \hline
   \end{tabular}
  \label{table4}  
\end{table}
Table 4 shows the derived temperature from the empirical relation. 
Uncertainty estimation was performed using the analytical law of propagation of errors. We accounted for two distinct sources of error, which are measurement uncertainties in pixel units ($\sigma_A$ and $\sigma_R$) and statistical uncertainties in the empirical model coefficients. These error components were combined in quadrature to derive the final uncertainty in the umbral temperature on Table 4.
The temperature of the sunspot umbra estimated from the umbral spectrum ranges around 4000 K, as shown in Section 3.
Most of the values in Table 4 are consistent with our results, except for the temperature derived from the Mathew's mean model.
This discrepancy between core or minimum(min) and mean temperature is not unexpected and can be explained by the fact that these temperatures represent different physical quantities.
The core and minimum umbral temperature is derived from the intensity measured in the darkest central region of the umbral. This measurement represents the thermal characteristics of the umbral core while minimizing the contributions from the brighter peripheral regions and represents the coolest detectable component of the umbra in the given dataset.
In contrast, the mean umbral temperature is derived from the averaged intensity over the entire umbra, including brighter regions around the umbral core.
Taken together, the consistency between the spectroscopic effective temperature and the empirically derived minimum and core temperatures suggests that the observed spectrum is best characterized by thermal conditions representative of the cooler components within the umbral components.
\begin{figure*}[h]
 \includegraphics[width=18cm]{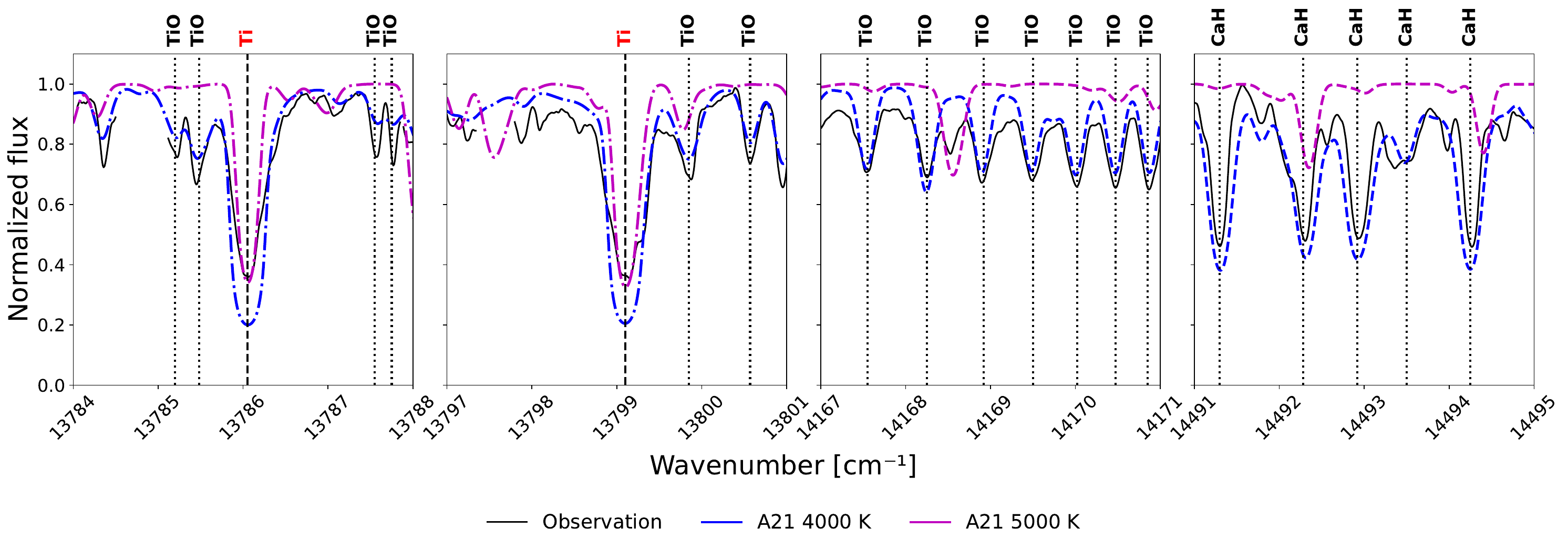} 
 \caption{Temperature sensitivity of atomic (Ti) and molecular (CaH and TiO) features compared with sunspot umbral observations. The Ti lines were selected from wavelength regions specifically chosen to avoid blending with molecular features. Two simulated spectra are shown: a 5000 K model (temperature close to cool model \citep{Komori_2025} and a 4000 K model (representing the best fit for molecular bands in our study). Both models assumed A21 \citep{Asplund2021} solar abundances and utilize a comprehensive line list, including line list $\#$1 and other species data from cool7.iso.lst. 
 The vertical dashed and dotted lines show the known line positions from \citep{Wallace1999AnAO}.
(Left Panels) While the 5000 K model reproduces the depth of the Ti I lines more effectively, it fails to account for the prominent molecular absorption features. (Right Panels) Conversely, the molecular band-heads are accurately fitted by the 4000 K model, whereas the 5000 K model predicts significantly weaker absorption than observed. This discrepancy highlights the inherent difficulty in reconciling atomic and molecular indicators within a single 1D-LTE framework.
\\{Alt text: A four panel spectrum comparison plot of sunspot umbral data against 4000 K and 5000 K synthetic models. The left panels show that while a 5000 K model fits atomic Ti I lines, it lacks molecular features. The right panels show that a 4000 K model accurately fits molecular CaH and TiO bands, while the 5000 K model fails to reproduce the molecular absorption depth.}}
\label{fig2}
\end{figure*}
\subsection{Contribution from other species}
As noted above, our spectral simulations yielded an effective temperature of around 4000 K, primarily because the CaH and TiO molecular features dominate the target wavelength range, whereas other atomic and molecular lines were not incorporated into the model. In contrast, \citet{Komori_2025} reported higher effective temperatures of 4767 K and 5170 K for the same sunspot umbra \citep{Wallace2000AnAO} over the wavenumber range of $\rm{15000\;cm^{-1}-25500\;cm^{-1}}$, and attributed this result to the limitations of 1D/Local thermodynamic equilibrium (LTE) spectral line formation and the need to include magnetic effects. This difference in the derived temperatures can be explained by the different spectral diagnostics emphasized in each study. \citet{Komori_2025} used a comprehensive line list containing more than 7000 lines from \citet{Brewer_2016}, which was calibrated against the quiet-Sun spectrum. As noted by \citet{Komori_2025}, however, this list does not include some of the molecular lines that become prominent in cooler umbral environments, causing their model to treat these regions as a smooth continuum even though molecular absorption is present. Their result therefore suggests that the model is driven toward warmer temperatures by the atomic features, and it also highlights the difficulty of simultaneously reproducing both atomic lines and the numerous molecular lines that appear in cooler umbrae.\\
Similarly, in our analysis, we found that while some atomic lines may be better reproduced by a higher-temperature model above 4500 K, such a model fails to account for the strong molecular absorption bands observed in the umbra in Fig.~5. Conversely, our model successfully reproduces the molecular features at 4000 K, but this temperature would likely provide an inconsistent fit to the atomic line depths, as discussed by \citet{Komori_2025}. Furthermore, \citet{Wallace2000AnAO} reported that this spectral region contains a very large number of TiO lines, on the order of 5000 - 10000, suggesting that these dominant molecular lines are important for reproducing cool spectra. Our sunspot umbral spectrum also contains many CaH and TiO lines \citep{Wallace1999AnAO}, and the derived temperature of 4000 K likely reflects the dominance of these molecular features. In addition, these molecular lines are less affected by magnetic fields, i.e., Zeeman shifts and splitting \citep{Wallace1999AnAO} and can therefore be captured reasonably well by a simple cooler 1D/LTE model.\\
Taken together, these two studies consistently show that a simple cool solar model cannot simultaneously reproduce the observed atomic and molecular line profiles. This systematic discrepancy underscores the need to include more sophisticated physical effects, such as magnetic Zeeman shifts and splitting, changes in pressure broadening due to suppressed convection, and non-LTE effects, in order to obtain a physically realistic representation of sunspot atmospheres, as also noted by \citep{Smitha2021}. Nevertheless, our results suggest that although 1D/LTE models have inherent limitations, focusing on temperature-sensitive molecular features can still provide a more reliable estimate of the umbral thermal structure, provided that the line list is sufficiently complete to account for the dominant absorption sources.
\section{Conclusion}
We presented the assignments of the $A^2\Pi–X^2\Sigma^+$ electronic transitions of the CaH molecule in a sunspot umbral spectrum. Using PGOPHER simulations, we identified 224 new lines, including previously unidentified (3-3) lines and higher rotational transitions in the (0-0), (1-1), and (2-2) bands that were not included in the previous study \citep{Wallace1999AnAO}. From the perspective of molecular spectroscopy, high-resolution solar observations such as sunspot umbral spectra can reveal molecular transitions that are difficult to observe in laboratory environments, thereby contributing to the extension of molecular line lists, which serve as valuable reference data for the analysis and interpretation of astronomical spectra.\\
CaH is also an important molecular species in the spectroscopy of cool astronomical environments, including some exoplanet atmospheres. In particular, molecular species such as TiO and CaH are often discussed as potential opacity sources in high-temperature planetary atmospheres, and spectroscopic studies of CaH are expected to play an important role in interpreting exoplanet spectra and determining atmospheric composition.\\
Using a line list that includes these newly assigned transitions, we demonstrated that the observed sunspot umbral spectrum, dominated by CaH and TiO molecules, can be reproduced more accurately than with line lists based solely on the assignments from previous studies. Based on comparisons between the spectral simulations and the observed spectrum, we found that an effective temperature of around 4000 K best reproduces the sunspot umbral spectrum. This value is slightly lower than, or at the lower end of, the commonly cited range of sunspot umbral temperatures (approximately 3900 - 4800 K). In addition, using the umbral size, we estimated the brightness temperature of the sunspot umbra in this study. These results suggest that the observed CaH and TiO absorption features are more sensitive to the coolest molecular regions within the sunspot than to the spatially averaged temperature of the entire umbra. In other words, the observed spectrum likely reflects the physical state of the dark core, or the minimum-temperature region, of the umbra.
\section*{Supplementary data} 
The following supplementary data is available at PASJ online.
E-table 0-0, 1-1, 2-2, 3-3.
\begin{ack}
We thank Dr. Gray for providing CaH line list and cool7.iso.lst.  It was very useful to prepare our own line lists.  We acknowledge support from JSPS KAKENHI Grant No. 22 K03685.
This work was supported by the Solar Sketch Database of the Solar Science Observatory, National Astronomical Observatory of Japan (NAOJ).
\end{ack}
\appendix 
\onecolumn
\section{$\text{C}_2$ and $\text{CN}$ molecular lines in $\mathrm{14000\;cm^{-1}-15000\;cm^{-1}}$ }
In the spectral region analyzed in this study ($\mathrm{14000\;cm^{-1}-15000\;cm^{-1}}$), there are numerous lines of $\text{C}_2$ and $\text{CN}$. To evaluate whether the omission of these species significantly affects our synthetic spectra and subsequent temperature estimations, we computed isolated synthetic spectra for $\text{C}_2$ and $\text{CN}$ across the relevant temperature range (3500 K to 4750 K). As illustrated in Figure 6, the absorption strengths of $\text{C}_2$ and $\text{CN}$ are negligibly shallow compared to the dominant CaH and TiO features at characteristic sunspot temperatures.
\begin{figure}[H]
 \begin{center}
 \includegraphics[width=16cm]{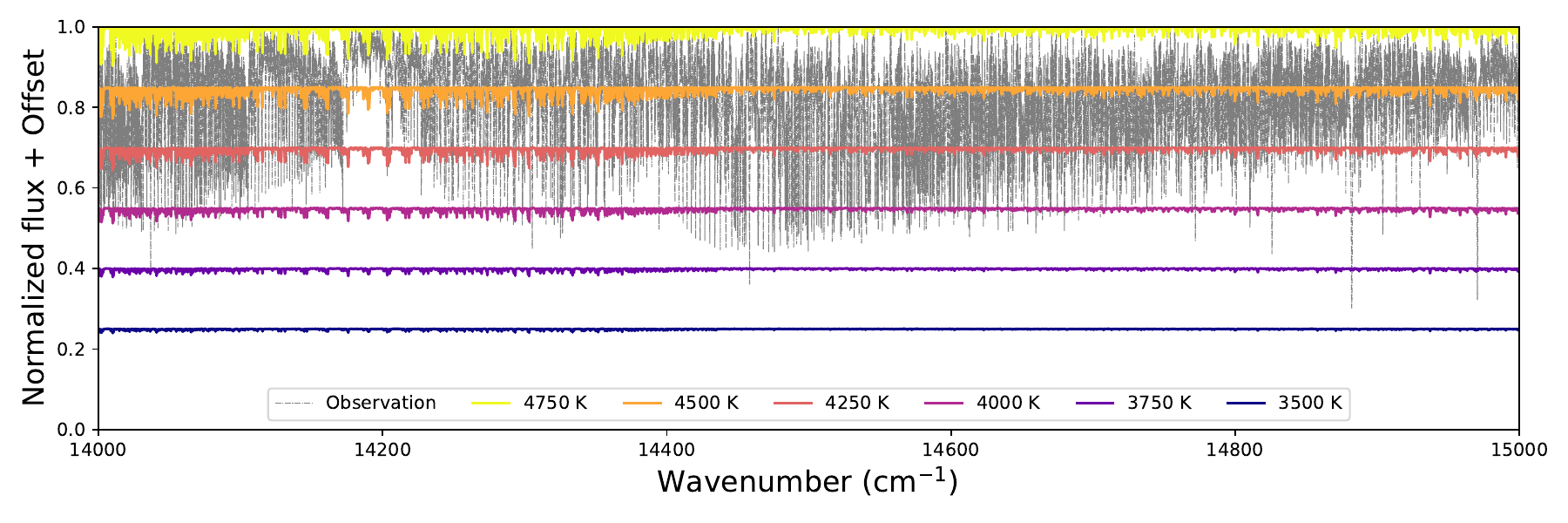} 
 \end{center}
 \caption{Comparison between the observed sunspot umbral spectrum and simulations including only $\text{C}_2$ and CN. While numerous $\text{C}_2$ and CN lines are present in this wavenumber region, their contributions are significantly smaller than those of CaH and TiO at sunspot temperatures. For clarity, the molecular spectra are vertically offset. \\{Alt text: Spectral comparison between observed data and calculations at 3500, 3750, 4000, 4250, 4500 and 4750 Kelvin using line list including only $\text{C}_2$ and CN molecules. The figure displays the less contributions of $\text{C}_2$ and CN lines within our target wavenumber region. } }
\label{fig2}
\end{figure}
\section{Consideration of other species and abundances}
To ensure the robustness of our derived effective temperature ($\sim$4000 K), we investigated the sensitivity of the spectral fitting to the inclusion of additional atomic and molecular species and different reference solar abundances. The left panel of Figure 7 compares the sum of squared residuals using line list (line list $\#$1) against lists supplemented with $\text{C}_2$, CN and other species from the cool7.iso.lst. In all cases, the minimum residual consistently points to 4000 K. Similarly, the right panel of Figure 7 demonstrates that the effect of  adopting different reference solar abundances (M22 (star) \citep{Magg2022}, A21 (circle) \citep{Asplund2021}, G96 (square) \citep{Grevesse1996} and G73 (right triangle) \citep{Gaur1972}). As shown in the figure, the three most recent abundance sets (M22, A21 and G96) give consistent results, all pointing to an optimal temperature minimum at 4000 K. While adopting the older values from \citet{Gaur1972} alters the absolute residual values, it also leads to similar estimates. 
\begin{figure}[H]
  \begin{center}
    \begin{minipage}{0.48\textwidth}
      \centering
      \includegraphics[width=.9\linewidth]{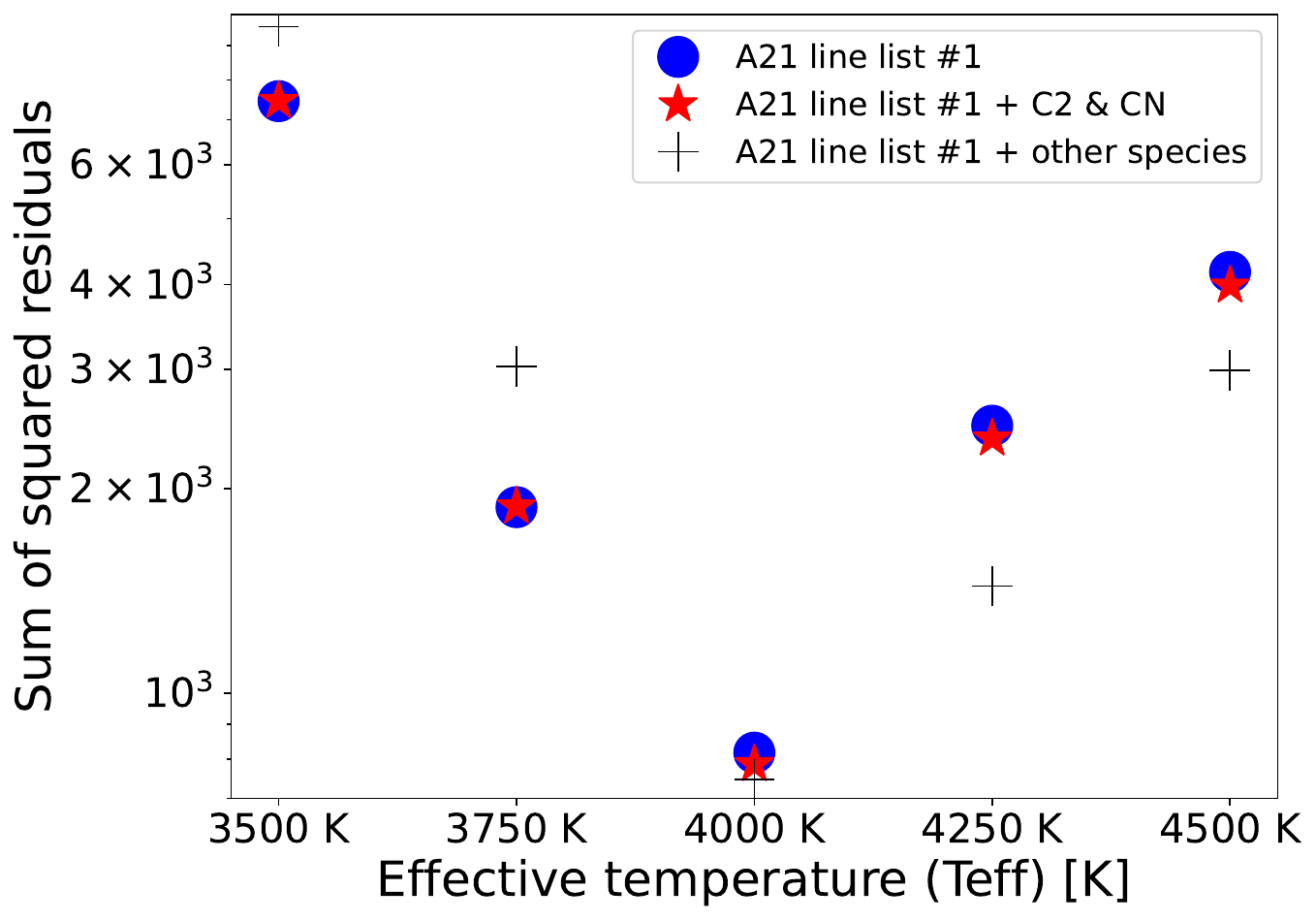}
    \end{minipage}
    \hfill
    \begin{minipage}{0.48\textwidth}
      \centering
      \includegraphics[width=.9\linewidth]{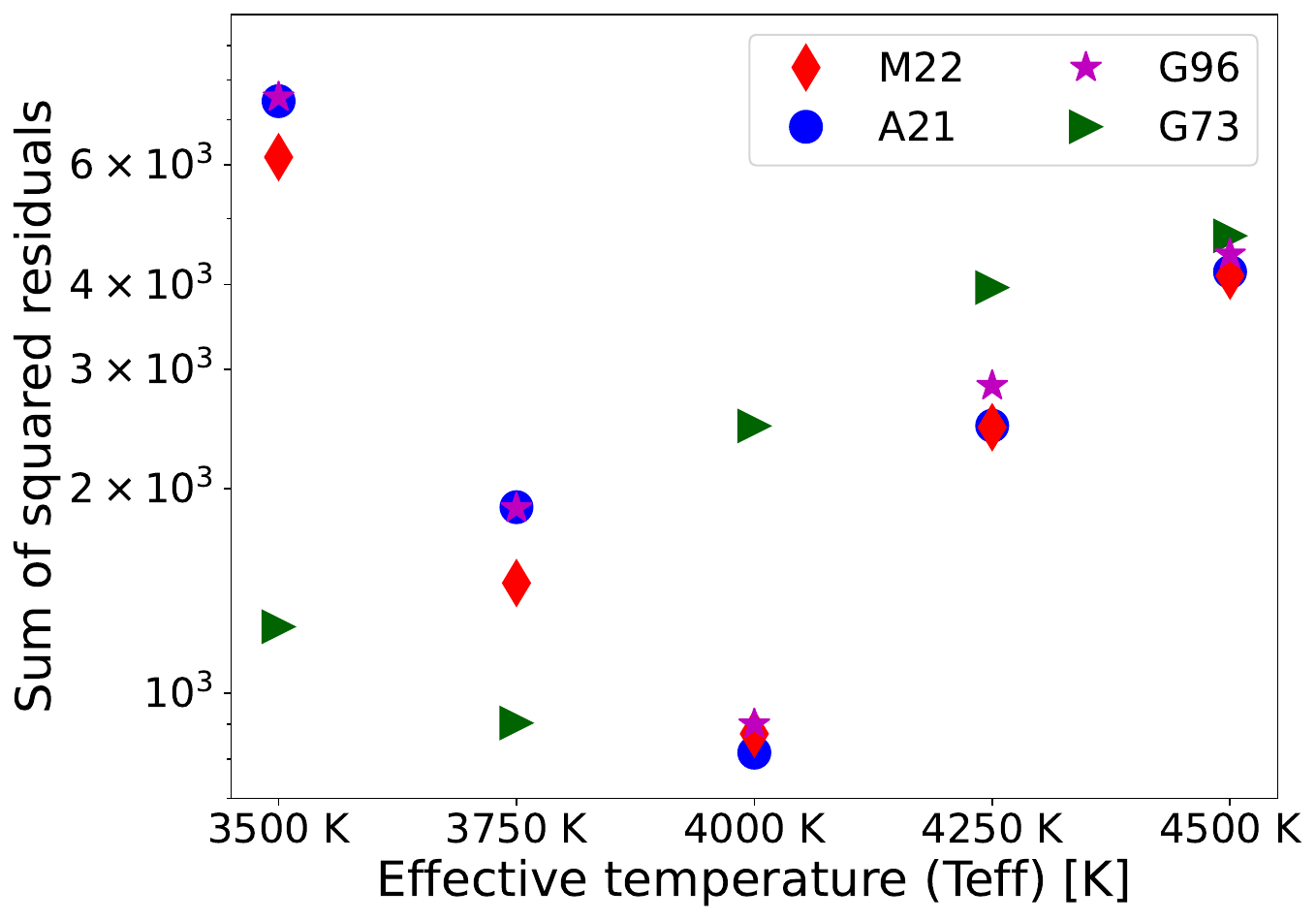}
    \end{minipage}
  \end{center}
  \caption{Temperature versus the sum of squared residuals for assumed different line list and abundances cases. Results are shown for temperatures of 3500, 3750, 4000, 4250, and 4500 K. Left : the result with A21 \citep{Asplund2021} abundance and three line lists including line list $\#$1 (circle), line list $\#$1 $+$ {$\text{C}_2$ and CN} (star) and line list $\#$1 $+$ {other atoms and molecules} from the cool7.iso.lst (plus). All three line list cases minimized the sum of squared residuals in the $4000 \text{ K}$.   Right : The result with four different abundances, M22 (star) \citep{Magg2022}, A21 (circle) \citep{Asplund2021}, G96 (square) \citep{Grevesse1996} and G73 (right triangle) \citep{Gaur1972}. 
  In both panels, although abundance uncertainties were evaluated as Figure 2, only the results using central values are plotted here for clarity. \\{Alt text: Scatter plot showing how the sum of squared residuals changes with temperature from 3500 to 4500 Kelvin depending on additional atomic and molecular species and different abundances. It demonstrates the sensitivity of the residuals to the choice of abundance and line list for each temperature point.}}
  \label{fig:spectra_3500}
\end{figure}
\section{Simulation using line list $\#$1 and A21 at different temperature}
Section 3.2 quantitatively demonstrates that the umbral spectrum is best reproduced at an effective temperature of 4000 K based on the sum of squared residuals (Figure 2). To provide visual confirmation of this behavior, Figures 8 and 9 display the comparison between the observed umbral spectrum and the calculated spectra using line list $\#$1 and A21 abundance \citep{Asplund2021} at off-optimal temperatures (3500 K, 3750 K, 4250 K, and 4500 K). At lower temperatures (<4000 K), the synthetic molecular absorption features become too deep, whereas at higher temperatures (>4000 K), the molecular bands significantly weaken and fail to match the observed umbral depths. These visual comparisons support the conclusion that 4000 K is the most appropriate 1D/LTE effective temperature for the molecular line-forming region in this specific sunspot umbra.
\begin{figure}[H]
\begin{center}
    \begin{minipage}{0.48\textwidth}
      \centering
      \includegraphics[width=.9\linewidth]{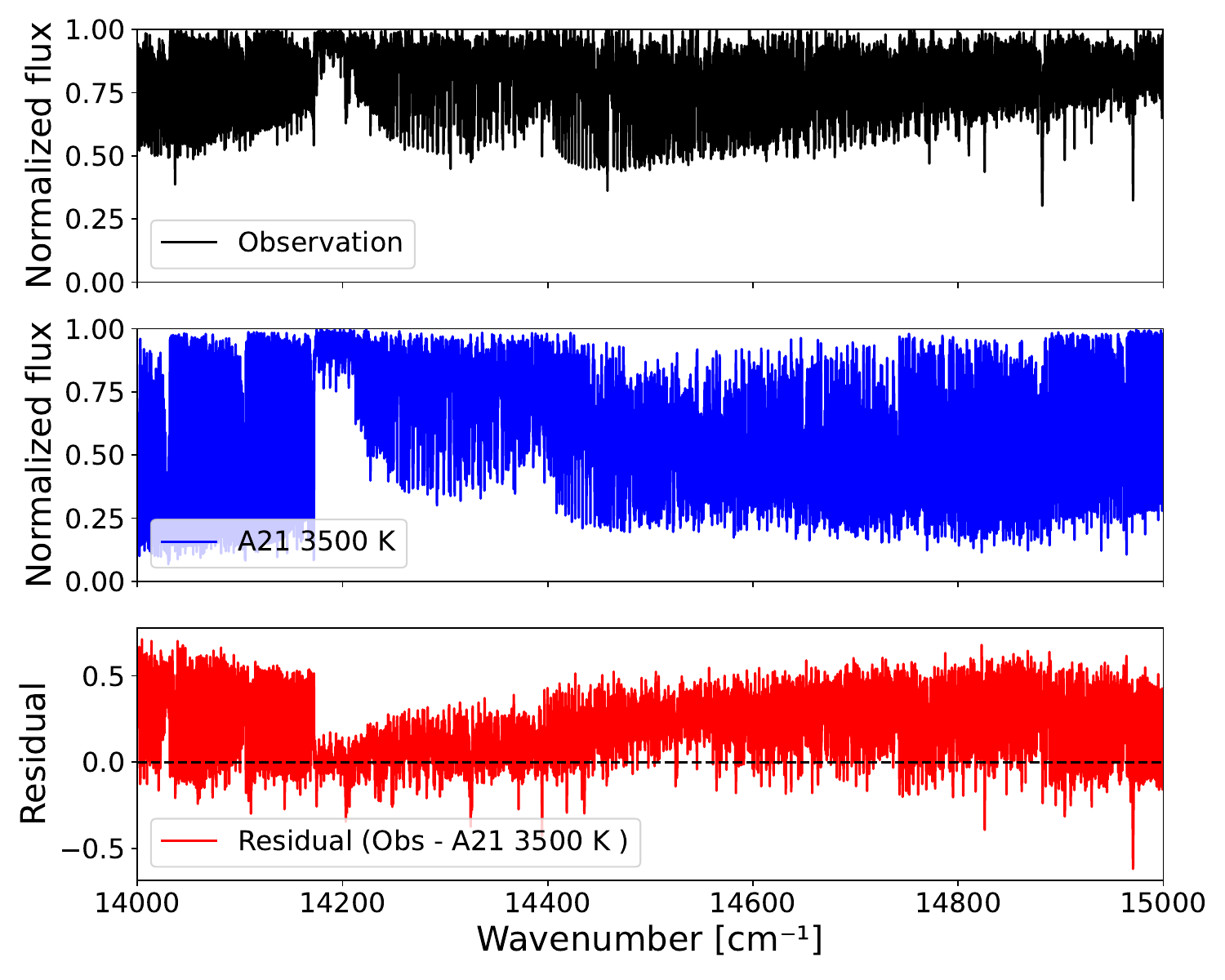}
    \end{minipage}
    \hfill 
    \begin{minipage}{0.48\textwidth}
      \centering
      \includegraphics[width=.9\linewidth]{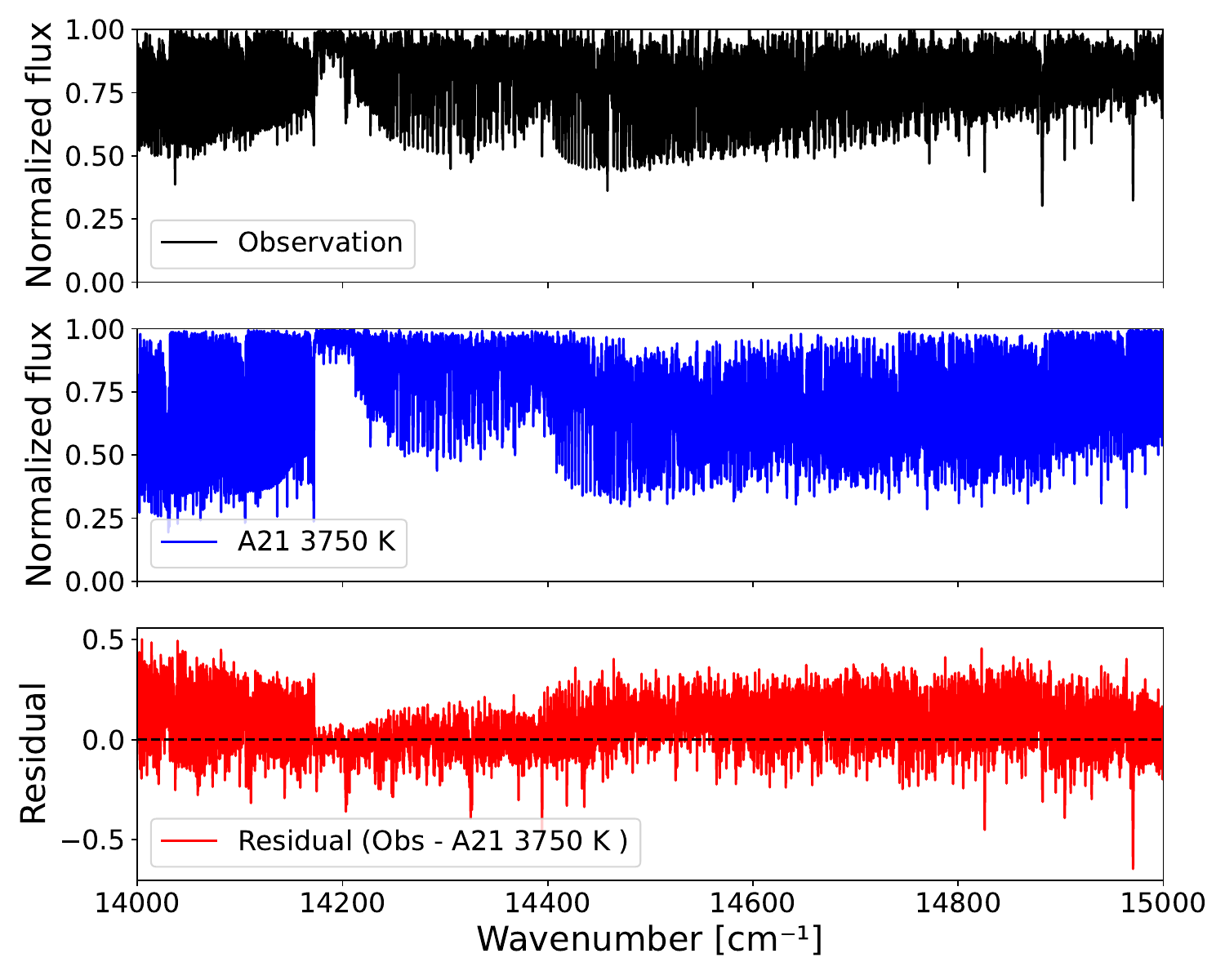}
    \end{minipage}
  \end{center}
  \caption{Comparison of the observed and calculated spectra (line list $\#$1) at 3500 K (left) and 3750 K (right).  \\{Alt text: Spectral comparison between observed data and calculations at 3500 and 3750 Kelvin using line list number 1. The figure displays the performance of two different temperatures in reproducing the observed spectrum.}}
  \label{fig:spectra_3500}
\end{figure}
\begin{figure}[H]
  \begin{center}
    \begin{minipage}{0.48\textwidth}
      \centering
      \includegraphics[width=.9\linewidth]{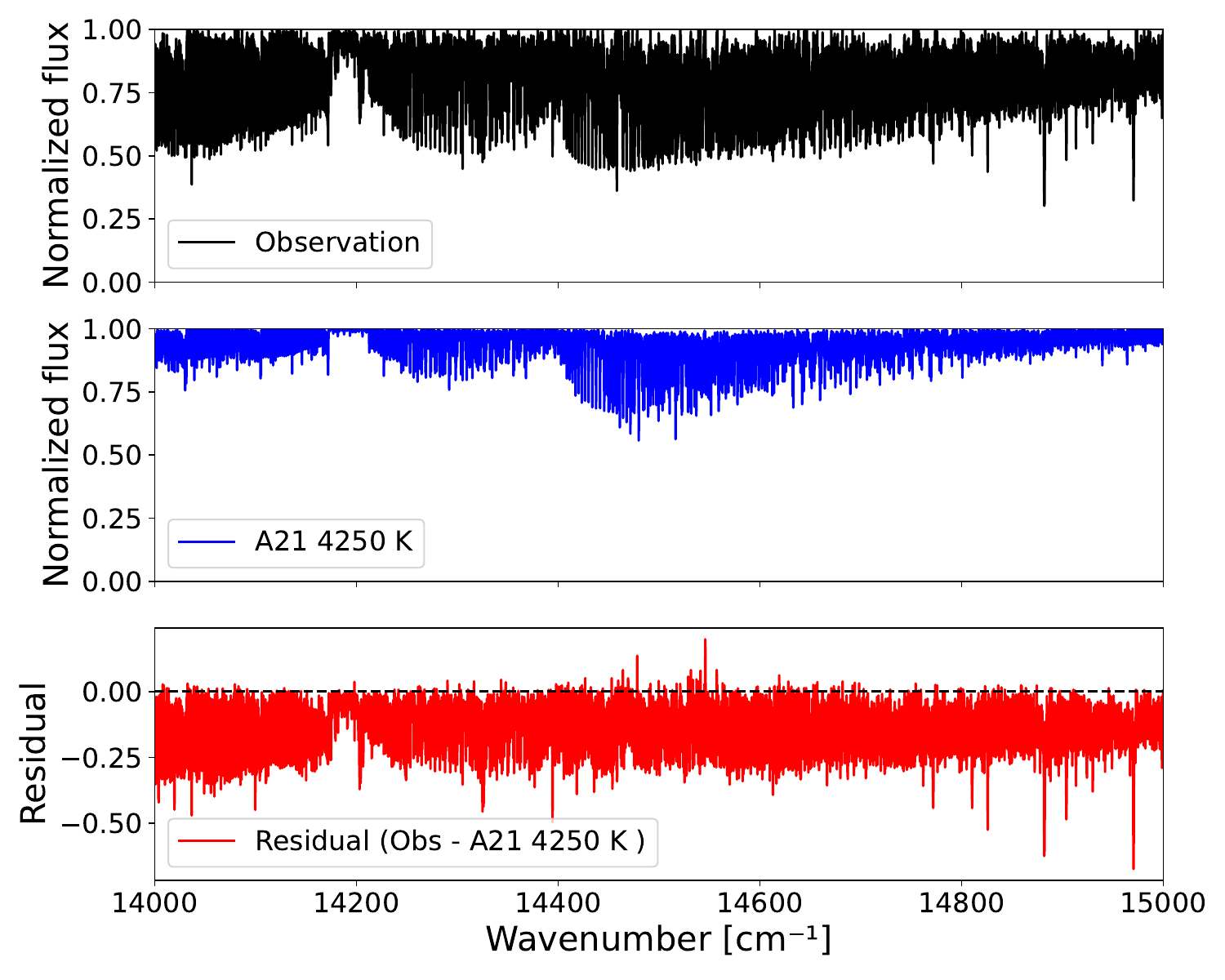}
    \end{minipage}
    \hfill
    \begin{minipage}{0.48\textwidth}
      \centering
      \includegraphics[width=.9\linewidth]{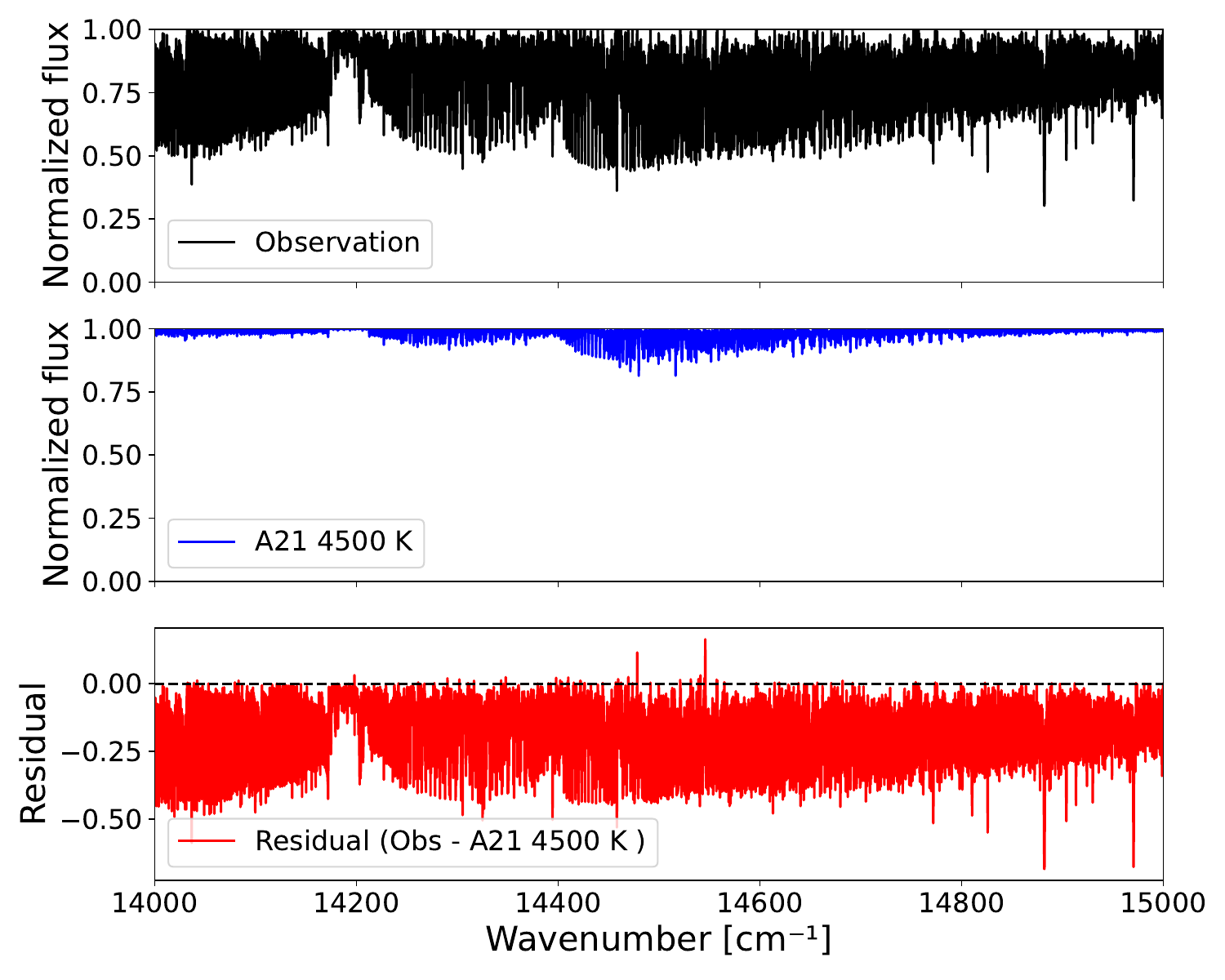}
    \end{minipage}
  \end{center}
  \caption{  Comparison of the observed and calculated spectra (line list $\#$1) at 4250 K (left) and 4500 K (right). \\{Alt text: Spectral comparison between observed data and calculations at 4250 and 4500 Kelvin using line list number 1. The figure displays the performance of two different temperatures in reproducing the observed spectrum.}}
  \label{fig:spectra_3500}
\end{figure}
\twocolumn
\bibliographystyle{apj}
\bibliography{ref_revised,Kit} 
\end{document}